\documentstyle[aps,prb,multicol,epsf]{revtex}

\begin{document}

\title{Analytical ground state for the three-band Hubbard model}

\author{C.~Waidacher, J.~Richter, R.~E.~Hetzel, and K.~W.~Becker}

\address{Institut f\"ur Theoretische Physik,
  Technische Universit\"at Dresden, D-01062 Dresden, Germany}
  
\maketitle

\begin{abstract}
For the calculation of charge excitations as those observed in, e.g.,
photo-emission spectroscopy or in electron-energy loss spectroscopy, a
correct description of ground-state charge properties is essential. In
strongly correlated systems like the undoped cuprates this is a 
highly non-trivial problem. In this paper we derive a non-perturbative 
analytical approximation for the ground state of the three-band Hubbard 
model on an infinite, half filled CuO$_{2}$ plane. 
By comparison with Projector Quantum Monte Carlo calculations it is shown 
that the resulting expressions correctly describe the charge properties of 
the ground state. Relations to other approaches are discussed. 
The analytical ground state preserves size consistency and can be 
generalized for other geometries, while still being both easy to interpret 
and to evaluate. 
\end{abstract}

\begin{multicols}{2}


\section{Introduction}

The interest devoted to the three-band Hubbard model\cite{Emery87} 
is due to the fact that it describes the charge 
properties of a CuO$_{2}$ layer, like those found in the high-$T_{C}$ 
superconducting cuprates, while still being comparatively 
simple.\cite{Dagotto94}  The basic 
assumption which leads to the three-band Hubbard model is that the only 
relevant orbitals are the Cu $3d_{x^{2}-y^{2}}$ and the O $2p_{x}$ and 
$2p_{y}$ orbitals. This granted, the CuO$_{2}$ layer may be described by a 
lattice with one Cu and two O sites per unit cell with hybridization between 
nearest neighbor Cu-O pairs and O-O pairs. In the hole picture the three-band 
Hubbard Hamiltonian reads 
\begin{mathletters}
\begin{eqnarray}
H &=&H_{0}+H_{1}~\mbox{,}\label{3-band1}\\
H_{0} &=&\Delta\sum_{j\sigma}n^p_{j\sigma}
+U_{d} \sum_{i}n^d_{i\uparrow}n^d_{i\downarrow}~\mbox{,}\label{3-band2}\\
H_{1} &=&t_{pd}\sum_{\langle ij\rangle \sigma}\phi^{ij}_{pd}
(p^\dagger_{j\sigma}d_{i\sigma}+h.c.) \nonumber \\
&+&t_{pp}\sum_{\langle jj^\prime\rangle \sigma}\phi^{jj^\prime}_{pp}
p^\dagger_{j\sigma}p_{j^\prime\sigma}~\mbox{,}\label{3-band3}
\end{eqnarray}
\end{mathletters}
where $d^\dagger_{i\sigma}$ ($p^\dagger_{j\sigma}$) create a hole 
with spin $\sigma$ in the $i$-th Cu $3d$ orbital ($j$-th O $2p$ 
orbital), while $n^d_{i\sigma}$ ($n^p_{j\sigma}$) are the 
corresponding number operators. $H_{0}$ is the atomic part of the 
Hamiltonian with the charge-transfer energy $\Delta$ and the on-site 
Coulomb repulsion $U_{d}$ between Cu $3d$ holes. $H_{1}$ represents 
the hybridization of Cu $3d$ and O $2p$ orbitals (hopping strength 
$t_{pd}$) and of O $2p$ orbitals (hopping strength $t_{pp}$). The 
factors $\phi^{ij}_{pd}$ and $\phi^{jj^\prime}_{pp}$ give the correct 
sign for the hopping processes\cite{Dagotto94}, and $\langle ij\rangle$ 
denotes the summation over nearest neighbor pairs.

In Eq.~(\ref{3-band1}) only the most important Coulomb repulsion $U_{d}$ is
included, while O on-site and inter-site Coulomb repulsions have been
neglected. For explicit calculations the following typical set of values 
for the parameters involved in Eq.~(\ref{3-band1}) will be 
used\cite{McMahan88}
\begin{eqnarray}
\Delta&=&3.5\text{ eV, }U_{d}=8.8\text{ eV,} \nonumber \\
t_{pd}&=&1.3\text{ eV, }t_{pp}=0.65\text{ eV.} \label{parameterset}
\end{eqnarray}
Strong correlations due to $U_{d}$ are the reason why ground state
properties of Hamiltonian $H$ can be calculated only
approximately and/or on finite clusters. Besides analytical approaches like,
e.g., different approximations for dynamical Green's functions\cite{Oles89}
mostly numerical simulations\cite{Stephan89} have been applied to the 
three-band Hubbard model.
The aim of the present work is to derive an analytical approximation for the
ground state of an infinite system at half-filling (i.e. one hole per Cu
site) which correctly describes charge properties and is still
comparatively easy to evaluate. The resulting approximation does not only
allow for a calculation of ground-state properties of the three-band model 
(\ref{3-band1}). It also provides a framework for the investigation of
excitations. Furthermore, the approach is sufficiently general to be applied
not only to a CuO$_{2}$ plane but also to different geometries like, e.g.,
that of a CuO$_{3}$ corner-sharing chain.\cite{Waidacher99}

Starting point of the approximation is a N\'{e}el-ordered ground state of
the atomic Hamiltonian $H_{0}$, which is denoted by $\left|
\psi _{0}\right\rangle $. Due to fluctuations induced by $H_{1}$ this atomic
ground state $\left| \psi _{0}\right\rangle $ differs from the full ground
state $\left| \Psi \right\rangle $ of Hamiltonian $H$. In Sec.~III it will
be shown that a perturbative treatment of these fluctuations breaks down for
parameter values which are in the physically relevant range. Therefore,
ground-state fluctuations have to be treated in a non-perturbative way. In
the following we will present a systematic and non-perturbative scheme to
introduce these fluctuations on the background of $\left| \psi
_{0}\right\rangle $.

The paper is organized as follows. In Sec.~II the general formalism is
presented. As an illustration in Sec.~III this formalism is applied to the
(exactly solvable) problem of a single CuO$_{4}$ plaquette. The
approximative ground state of an infinite, half-filled CuO$_{2}$ plane is
developed in Sec.~IV, and ground-state expectation values are evaluated in
Sec.~V. In Sec.~VI the results of the analytical approach are compared to
Projector Quantum Monte Carlo simulations. The conclusions are presented
in Sec.~VII. Finally, a more detailed justification of the approach 
together with a discussion of\ its relationship to the cumulant formalism 
\cite{Becker88}\ is given in the Appendix.


\section{General formalism}

The basic idea is to start with a state $\left| \psi _{0}\right\rangle $
which is a first approximation to the full ground state $\left| \Psi
\right\rangle $. In the present case we will choose $\left| \psi
_{0}\right\rangle $ to be a N\'{e}el-ordered ground state of the atomic
Hamiltonian $H_{0}$, Eq.~(\ref{3-band2}). In $\left| \psi _{0}\right\rangle $ 
every Cu site is singly occupied and all O sites are empty. Next, 
fluctuations on the background of $\left| \psi _{0}\right\rangle $ are 
introduced. These ground-state fluctuations are described by fluctuation 
operators $F_{\alpha }$ (introduced below in more detail) which approximately 
transform state $\left| \psi _{0}\right\rangle $ into the full ground state 
$\left| \Psi\right\rangle $. Under quite general conditions it can be shown 
\cite{Schork92} that the transformation leading from 
$\left| \psi_{0}\right\rangle $ to $\left| \Psi \right\rangle $ 
has to be of exponential form 
\begin{equation}
\left| \Psi \right\rangle =\exp \left( \sum_{\alpha }\lambda _{\alpha
}F_{\alpha }\right) \left| \psi _{0}\right\rangle \text{ .}
\label{groundstate1}
\end{equation}
The parameters $\lambda _{\alpha }$ are fluctuation strengths of the
fluctuation operators $F_{\alpha }$. They are determined using
the set of equations 
\begin{equation}
0=\left\langle \Psi \right| \left[ H,F_{\alpha }^{\dagger }\right] \left|
\Psi \right\rangle \text{ , }\alpha =1,2,\ldots\quad\mbox{.}  
\label{lambda-equations1}
\end{equation}
Eq.~(\ref{lambda-equations1}) follows from the condition that $\left| \Psi
\right\rangle $ is an eigenstate of the full Hamiltonian $H$, 
Eq.~(\ref{3-band1}). From Eq.~(\ref{groundstate1}) all ground-state properties 
can be evaluated using 
\begin{equation}
\left\langle A\right\rangle =\frac{\left\langle \Psi \right| A\left| \Psi
\right\rangle }{\left\langle \Psi \mid \Psi \right\rangle }~\mbox{.}
\label{expectationvalue}
\end{equation}
The ground-state energy, for instance, is calculated from Eq.(\ref
{expectationvalue}) with $A=H$. Equations~(\ref{groundstate1}) and (\ref
{lambda-equations1}), together with an appropriate choice of fluctuation
operators $F_{\alpha }$, constitute the formal framework to be used in the
remainder of this work. The above formalism will be applied to an
exactly solvable problem in Sec.~III. This serves both as an illustration
of the approach, and as an indication to which extent a perturbative
treatment of ground-state fluctuations is possible.

We finally remark that Eqs.~(\ref{groundstate1}) and 
(\ref{lambda-equations1}) are closely related to the cumulant formalism
(see Appendix). Therefore the approach presented in this work preserves
size consistency. Thus, for instance, the approximated ground-state
energy remains an extensive quantity.


\section{Application to a single plaquette}

As an example, the method presented in the last section is now used to find
an approximate expression for the ground state of Hamiltonian~(\ref{3-band1})
for the case of a single CuO$_{4}$ plaquette occupied by a single hole. In
this case the unperturbed ground state $\left| \psi _{0}\right\rangle $ of 
$H_{0}$ is a state in which the Cu site is singly occupied while the four O
sites are empty. For reasons of symmetry we may use a single fluctuation
operator $F_{1}$ which describes fluctuations of the hole from
the Cu site to the four surrounding O sites $j$, i.e. 
\begin{equation}
F_{1}=-\sum_{j\sigma }\phi _{pd}^{j}p_{j\sigma }^{\dagger }d_{\sigma }~\mbox{,}
\label{plaqF1}
\end{equation}
where $\phi _{pd}^{j}$ are the phase factors introduced in Eq.~(\ref{3-band3}).
According to Eq.~(\ref{groundstate1}) the full ground state of a hole on a
single plaquette is expressed by 
\begin{equation}
\left| \Psi \right\rangle =\exp \left( \lambda _{1}F_{1}\right) \left| \psi
_{0}\right\rangle~\mbox{.}   \label{plaqground-state}
\end{equation}
The norm of this state is 
\begin{equation}
\left\langle \Psi \mid \Psi \right\rangle =1+4\lambda _{1}^{2}\text{ .}
\label{plaqnorm}
\end{equation}
The fluctuation strength $\lambda _{1}$ in Eq.~(\ref{plaqground-state}) is
determined from condition~(\ref{lambda-equations1}) 
\begin{equation}
0=\left\langle \Psi \right| \left[ H,F_{1}^{\dagger }\right] \left| \Psi
\right\rangle \text{ .}  \label{plaqlambdaequation1}
\end{equation}
Non-vanishing contributions in Eq.~(\ref{plaqlambdaequation1}) arise only 
from terms up to order $\lambda _{1}^{2}$. The following quadratic 
equation for $\lambda _{1}$ is obtained 
\begin{equation}
0=4t_{pd}-4\left( \Delta -2t_{pp}\right) \lambda _{1}-16t_{pd}\lambda
_{1}^{2}\text{ .}  \label{plaqlambdaequation2}
\end{equation}
When the positive solution for $\lambda _{1}$ is used in 
Eq.~(\ref{plaqground-state}) one obtains the exact ground state. The
ground-state energy $E_{G}$ is calculated using Eq.~(\ref{expectationvalue}) 
\begin{eqnarray}
E_{G}&=&-4t_{pd}\lambda _{1} \nonumber\\
&=&\frac{1}{2}\left[ \Delta -2t_{pp}-\sqrt{\left(
\Delta -2t_{pp}\right) ^{2}+\left( 4t_{pd}\right) ^{2}}\,\right] \text{ .}
\label{plaqenergy}
\end{eqnarray}
For parameter set~(\ref{parameterset}) a value of $\lambda _{1}=0.33$
results. Notice that Eq.~(\ref{plaqenergy}) contains only a reduced effective
charge-transfer energy $\Delta -2t_{pp}$. The Cu-occupation number 
$\left\langle n_{{\rm Cu}}\right\rangle =\left\langle \Psi 
\right| n_{d}\left| \Psi
\right\rangle \left\langle \Psi \mid \Psi \right\rangle ^{-1}$ is given by 
\begin{eqnarray}
\left\langle n_{{\rm Cu}}\right\rangle  &=&\frac{\left\langle \psi _{0}\right|
\exp \left( \lambda _{1}F_{1}^{\dagger }\right) n_{d}\exp \left( \lambda
_{1}F_{1}\right) \left| \psi _{0}\right\rangle }{\left\langle \Psi \mid \Psi
\right\rangle } \nonumber  \\
&=&\frac{\left\langle \psi _{0}\right| \left( 1+\lambda _{1}F_{1}^{\dagger
}\right) n_{d}\left( 1+\lambda _{1}F_{1}\right) \left| \psi
_{0}\right\rangle }{\left\langle \Psi \mid \Psi \right\rangle }  \nonumber \\
&=&\frac{1}{1+4\lambda _{1}^{2}}~\mbox{.}  \label{plaqncu}
\end{eqnarray}
From this result one may conclude that a perturbative treatment of $F_{1}$
fluctuations (i.e.\ an expansion in $\lambda _{1}$, see Appendix),
also in infinite systems, is in general not possible. 
Typically, $\lambda _{1}$ is of the order $1/2$. An expansion of 
Eq.~(\ref{plaqncu}) in $\lambda _{1}$, however, diverges for 
$\lambda_{1}\geq 0.5$. Condition $\lambda _{1}=0.5$ is equivalent to a 
vanishing effective charge-transfer energy $\Delta -2t_{pp}=0$, and to
a Cu-occupation number of $1/2$. 
At this point state $\left| \psi _{0}\right\rangle $ ceases to be a
good approximation of the exact ground state $\left| \Psi \right\rangle $.
This divergence has been observed previously\cite{Becker90}, although its
origin was unclear at that time.


\section{Application to an infinite CuO$_{2}$ plane}

We now apply the formalism presented in Sec.~II to the geometry of an
infinite CuO$_{2}$ plane. State $\left| \psi _{0}\right\rangle $ is again 
the N\'{e}el-ordered ground state of the atomic Hamiltonian $H_{0}$,
Eq.~(\ref{3-band2}). Let us introduce appropriate fluctuation 
operators $F_{\alpha}$. First, operator $F_{1}$ from Eq.~(\ref{plaqF1}) 
is generalized to all $N$ Cu sites $i$
\[F_{i,1}=-\sum_{j\sigma }\phi _{pd}^{ij}\,p_{j\sigma }^{\dagger }d_{i\sigma } 
~\mbox{,}\]
where the sum is over the four O sites $j$ which surround Cu site $i$. 
The remaining operators are constructed in accordance with the following
principles: (i) All operators describe delocalizations of a hole initially
located at Cu site $i$. (ii) The final site in the process is reached via
the shortest path accessible by Cu-O hopping processes. (iii) A summation
over all equivalent final sites is taken. (iv) The signs of the hopping
processes are chosen to be the negative of the phases $\phi _{pd}^{ij}$
in Hamiltonian~(\ref{3-band3})(which guarantees non-negative values for 
the fluctuation strengths $\lambda _{\alpha }$).

\epsfxsize=0.2\textwidth
\epsfbox{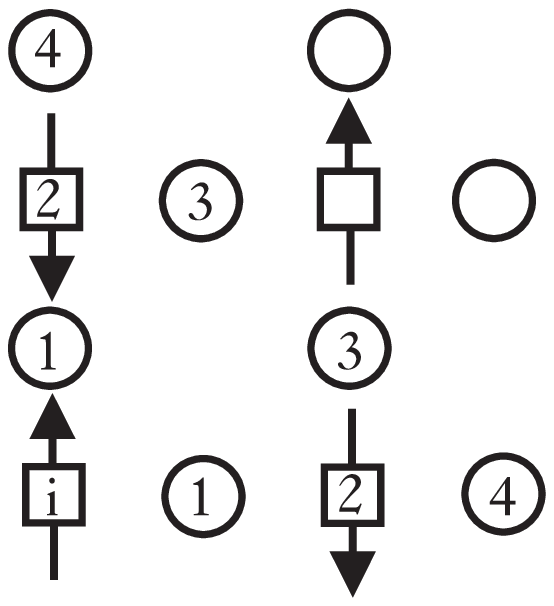}
\begin{figure}
\narrowtext
\caption{Final sites of the fluctuation operators $F_{i,\alpha }$. 
Cu and O sites are symbolized by squares and circles, respectively. 
Arrows show spin orientation and position of the holes in the atomic 
ground state. The fluctuations $F_{i,\alpha }$ start from Cu site $i$ 
and lead to the final sites sites labelled by $\alpha=1,\ldots 4$. 
The area shown is a quarter of the full area accessible to the fluctuations.}
\end{figure}

Figure~1 shows final sites reached by fluctuation operators $F_{i,\alpha }$, 
$\alpha =1,\ldots 4$. For reasons of symmetry only a quarter of the allowed
fluctuation range is shown. Fluctuation $F_{i,2}$, for instance,
describes the hopping of the hole from Cu site $i$, via O site $j$, 
to the four nearest-neighbor Cu sites $k$
\[
F_{i,2}=-\sum_{jk\sigma }\left( 1-n_{j
\sigma }^{p}\right) d_{k\sigma}^{\dagger }d_{i\sigma }~\mbox{.}
\]
Final states with singly or doubly occupied Cu sites differ by the Coulomb
energy $U_{d}$. Since $U_{d}$ is large we have to distinguish between these
two cases. Therefore we split $F_{i,2}$ into two operators which describe a
process leading to a singly or doubly occupied Cu site, respectively 
\begin{eqnarray*}
F_{i,2s} &=&-\sum_{jk\sigma }\left( 1-n_{k
\overline{\sigma }}^{d}\right) \left( 1-n_{j\sigma }^{p}\right)  
d_{k\sigma }^{\dagger } d_{i\sigma }~\mbox{,} \\
F_{i,2d} &=&-\sum_{jk\sigma }n_{k\overline{
\sigma }}^{d} \left( 1-n_{j\sigma }^{p}\right) d_{k\sigma }^{\dagger }
d_{i\sigma }~\mbox{.}
\end{eqnarray*}
Note that it is not necessary to introduce a fluctuation operator which
leads to the nearest neighbor Cu sites in diagonal direction (e.g. the Cu
site without a label in Fig.~1). Due to the Pauli principle fluctuations to
these sites are largely excluded because of antiferromagnetic order. The 
neglect of fluctuations leading beyond the range shown in Fig.~1 will be 
justified \textit{a posteriori}. It will be shown that the fluctuation 
strengths $\lambda _{\alpha }$ decrease rapidly with increasing length of 
the fluctuation processes.

According to Eq.~(\ref{groundstate1}) the ground state has the following form 
\begin{equation}
\left| \Psi \right\rangle =\exp \left( \sum_{i\alpha }\lambda _{\alpha
}F_{i\alpha }\right) \left| \psi _{0}\right\rangle~\mbox{,}  
\label{groundstate2}
\end{equation}
where $\alpha $ denotes the $5$ fluctuation operators described above.
Because of translational symmetry the parameters $\lambda _{\alpha }$ do not
depend on the Cu-site index $i$. To simplify Eq.~(\ref{groundstate2}) we
approximately factorize the exponential function with respect to the
fluctuations $F_{i,1}$
\begin{equation}
\left| \Psi \right\rangle =\exp \left( \sum_{i\alpha >1}\lambda _{\alpha
}F_{i\alpha }\right) \exp \left( \sum_{i^{\prime }}\lambda _{1}F_{i^{\prime
},1}\right) \left| \psi _{0}\right\rangle~\mbox{.}  \label{groundstate3}
\end{equation}
This approximation amounts to the assumption that far-reaching fluctuations $
F_{i,\alpha >1}$ occur on the background of $F_{i,1}$-fluctuations which in
turn are influenced only indirectly (i.e. via $\lambda _{1}$) by the former.
The second exponential function in Eq.~(\ref{groundstate3}) exactly
factorizes with respect to $i^{\prime }$
\begin{equation}
\left| \Psi \right\rangle =\exp \left( \sum_{i\alpha >1}\lambda _{\alpha
}F_{i\alpha }\right) \prod_{i^{\prime }}\left( 1+\lambda _{1}F_{i^{\prime
},1}\right) \left| \psi _{0}\right\rangle~\mbox{.}  \label{groundstate4}
\end{equation}
In Eq.~(\ref{groundstate4}) every hole may fluctuate over a total 
range of five plaquettes each. Notice that all holes fluctuate 
simultaneously. This leads to a multitude of many-body effects, i.e. 
the fluctuation of a hole depends on the configuration of other holes.
Basically there are three types of many body effects which are
exemplified in Fig.~2.
First, due to the Pauli principle the fluctuation of a hole may be blocked
by the presence of other holes with the same spin, as in the fluctuation
process labelled (a) in Fig.~2. Second, there are processes in which 
holes with the same spin change place, cf. process (b). 
In the following we will call these processes \textit{site-changing
processes}. Third, there are strong correlations due to the Hubbard $U_{d}$
on doubly occupied Cu sites, as in process (c). One common feature of all 
these many-body effects is that they suppress fluctuations.

\epsfxsize=0.2\textwidth
\epsfbox{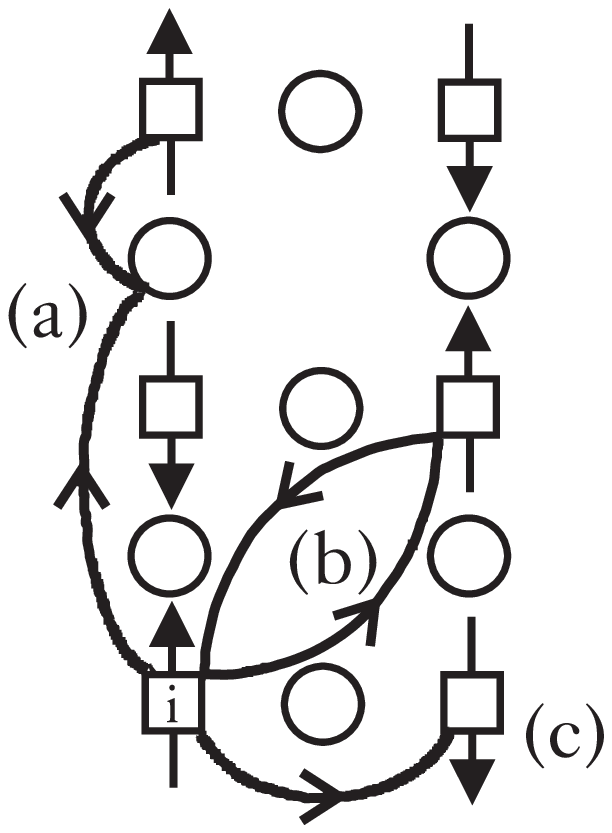}
\begin{figure}
\narrowtext
\caption{Examples for many body effects. There are three types of effects: 
(a) processes which are excluded by the Pauli-principle, (b) site-changing 
processes, and (c) correlations due to the Hubbard $U_{d}$ on doubly 
occupied Cu sites. The approach presented in this work accounts for all 
effects shown here.}
\end{figure}

This multitude of many-body effects makes an exact evaluation of 
expectation values using Eq.~(\ref{groundstate4}) impossible. 
Further approximations are therefore necessary. Let us consider 
processes in which two or more holes simultaneously leave their 
original plaquette. The fluctuation strengths $\lambda _{\alpha}$ 
for such far-reaching fluctuations turn out to be small compared to 
$\lambda _{1}$. Therefore it should be possible to neglect the many-body
effects arising in these processes (except for site-changing processes
in diagonal direction, see below). In the case of $F_{i,2d}$ fluctuations, 
for example, we neglect the possibility that the O site $j$ between the 
Cu starting and final sites $i$ and $k$ may already be occupied by a hole 
with the same spin. This amounts to the following simplification
\[
F_{i,2d} =-\sum_{jk\sigma }n_{k\overline{
\sigma }}^{d}~d_{k\sigma }^{\dagger }d_{i\sigma }~\mbox{.} \nonumber
\]
In this way all of the aforementioned
processes are included, some of them however only in a simplified way
(i.e.\ by neglecting many-body effects).

In addition to all many-body effects which are due to processes where only 
one hole leaves
its original plaquette, we furthermore take account of all site-changing
processes in diagonal direction, see Fig.~2(b). The suppression of charge
fluctuations due to the diagonal sites turns out to be of great importance.
On the other hand, site-changing processes involving next-nearest Cu
neighbors in horizontal or vertical direction can be neglected for the
following reason: During these processes the paths of the holes have to
cross at the intermediate O sites (i.e. sites 1 and 4 in Fig.~1). However,
since the holes have the same spin, they have to avoid each other due to the
Pauli principle. Site-changing processes in horizontal or vertical direction
are therefore unlikely.


\section{Evaluation of expectation values}

We now evaluate expectation values with state~(\ref{groundstate4}). Using the 
above approximations the norm of this state is $\left\langle \Psi \mid \Psi
\right\rangle =\nu ^{N}$, where $N$ is the number of Cu sites and, in generalization of Eq.~(\ref{plaqnorm})
\begin{equation}
\nu =1+\sum_{\alpha }z_{\alpha }p_{\alpha }\lambda _{\alpha }^{2}~\mbox{.}
\label{norm}
\end{equation}
$z_{\alpha }$ is the number of equivalent final sites of the given process
(e.g.\ $z_{2s}=4$), and $p_{\alpha }$ is the probability that the
configuration of the other holes makes the process possible. This
probability is defined by 
\begin{equation}
p_{\alpha }=\frac{\left\langle \Psi \right| P_{i,\alpha }\left| \Psi
\right\rangle }{\left\langle \Psi \mid \Psi \right\rangle }~\mbox{,}
\label{probabilities}
\end{equation}
where $P_{i,\alpha }$ is a projection operator on all configurations which
allow for process $\alpha $. For example, $P_{i,2s}$ projects on states in
which the target Cu site of fluctuation $F_{i,2s}$ is empty, whereas 
$P_{i,2d}$ is the projector on states with a singly occupied final site. 
Due to translational symmetry the probabilities $p_{\alpha }$ do not depend 
on the site index $i$. Obviously, Eqs.~(\ref{norm}) and (\ref{probabilities}) 
have to be solved self-consistently since $\left| \Psi \right\rangle$ 
in Eq.(\ref{probabilities}) depends on the parameters $\lambda _{\alpha}$. 
In the case of $p_{2s}$, for instance, we obtain by explicit calculation
\begin{eqnarray}
p_{2s} &=&\frac{\left\langle \Psi \right| P_{i,2s}\left| \Psi \right\rangle }
{\nu ^{N}} \nonumber\\
&=&\frac{1}{\nu }\sum_{\alpha }z_{\alpha }p_{\alpha }\lambda _{\alpha }^{2}
\nonumber\\
&=&1-1/\nu~\mbox{.} \nonumber
\end{eqnarray}
In an analogous way one finds 
\begin{eqnarray}
p_{1} &=&1~\mbox{,}~~p_{2s}=1-1/\nu~\mbox{,}~~p_{2d}=1/\nu~\mbox{,}   
\nonumber \\
p_{3} &=&1-2\lambda_{1}^{2}/\nu~\mbox{,}~~ 
p_{4}=1-\lambda _{1}^{2}/\nu~\mbox{.}  \label{explicitprobabilities}
\end{eqnarray}
The interpretation of Eq.~(\ref{explicitprobabilities}) is straightforward. 
$p_{1}=1$ holds since we assume that far-reaching fluctuations occur on the
background of $F_{i,1}$-fluctuations. $1/\nu $ is the probability to find a
given hole at its original Cu site. $p_{2d}$ is therefore the probability
that a target Cu site is singly occupied. This is a necessary prerequisite
for the fluctuation process $F_{2d}$ which leads to a double occupancy, cf.
Fig.~2(c). $p_{2s}$, on the other hand, is the probability that a target Cu
site is empty, as required for fluctuation process $F_{2s}$. The probability
to find a given hole at a specific O site on its original plaquette is 
$\lambda _{1}^{2}/\nu $. Thus $p_{4}$ is the probability that the target O
site of fluctuation $F_{4}$ is not blocked by the hole of same spin which
resides on the neighboring Cu site, cf. Fig.~2(a). Analogously, $p_{3}$ is the
probability that the target O site of fluctuation $F_{3}$ is not blocked.
The additional factor $2$ in $p_{3}$ (as compared to $p_{4}$) is due to
site-changing processes, cf. Fig.~2(b).

The fluctuation strengths $\lambda _{\alpha }$ are calculated using Eq.~(\ref
{lambda-equations1}) for an arbitrary site $i=0$ 
\begin{equation}
0=\left\langle \Psi \right| \left[ H,F_{0,\alpha }^{\dagger }\right] \left|
\Psi \right\rangle~\mbox{.}  \label{lambda-equations2}
\end{equation}
One obtains the following nonlinear system of equations
\begin{eqnarray}
0&=& \left(E_{G}-\Delta+2t_{pp}\right)\lambda_{1}+t_{pd}
+t_{pd}\lambda_{2s}p_{2s} \nonumber \\
&&+t_{pd}\lambda_{2d}p_{2d}+2t_{pp}\lambda_{3}p_{3}
-2t_{pp}\lambda _{1}^{2}\lambda _{3}/\nu~\mbox{,} \label{system1}  \\
0&=&E_{G}\lambda _{2s}+4t_{pd}\lambda _{2d}\lambda _{1}/\nu 
+t_{pd}\lambda _{1}p_{2s} \nonumber \\
&&+2t_{pd}\lambda _{3}p_{3}p_{2s}+t_{pd}\lambda _{4}p_{4}p_{2s}~\mbox{,} 
\label{system2}  \\
0&=&\left( 2E_{G}-U_{dd}\right)\lambda _{2d} 
+4t_{pd}\lambda _{2s}\lambda_{1} \nonumber \\
&&+t_{pd}\lambda _{1}+2t_{pd}\lambda _{3}p_{3}
+t_{pd}\lambda _{4}p_{4}~\mbox{,}\label{system3} \\
0&=&\left( E_{G}-\Delta +t_{pp}\right)\lambda_{3}p_{3} 
+t_{pp}\lambda_{1}p_{3} \nonumber \\
&&+t_{pd}\lambda _{2s}p_{2s}p_{3}
+t_{pd}\lambda _{2d}p_{2d}p_{3} \nonumber \\
&&+t_{pp}\lambda _{4}p_{3}
-\left( t_{pd}+2t_{pp}\lambda _{1}\right) 
\lambda _{1}\lambda _{3}/\nu~\mbox{,} \label{system4} \\
0&=&\left( E_{G}-\Delta \right)\lambda _{4}p_{4} +t_{pd}\lambda
_{2s}p_{2s}p_{4}+t_{pd}\lambda _{2d}p_{2d}p_{4} \nonumber \\
&&+2t_{pp}\lambda_{3}p_{3}-\left( t_{pd}+2t_{pp}\lambda _{1}\right) \lambda
_{1}\lambda _{4}/ \nu ~\mbox{,} \label{system5}
\end{eqnarray}
where $E_{G}=-4t_{pd}\lambda _{1}$ is the ground-state energy per Cu site
(see Eq.~(\ref{physicalquantities1})). This system of equations, together 
with Eqs.~(\ref{norm}) and (\ref{explicitprobabilities}) can be solved 
self-consistently for all $\lambda _{\alpha }$, $p_{\alpha }$ and for $\nu$. 
The solution with the lowest value of $E_{G}$ is then used in 
Eq.~(\ref{groundstate4}).
In the case of $\lambda _{\alpha }=0$ for all $\alpha >1$ Eq.~(\ref{system1})
reduces to Eq.~(\ref{plaqlambdaequation2}) for the single plaquette.

\epsfxsize=0.4\textwidth
\epsfbox{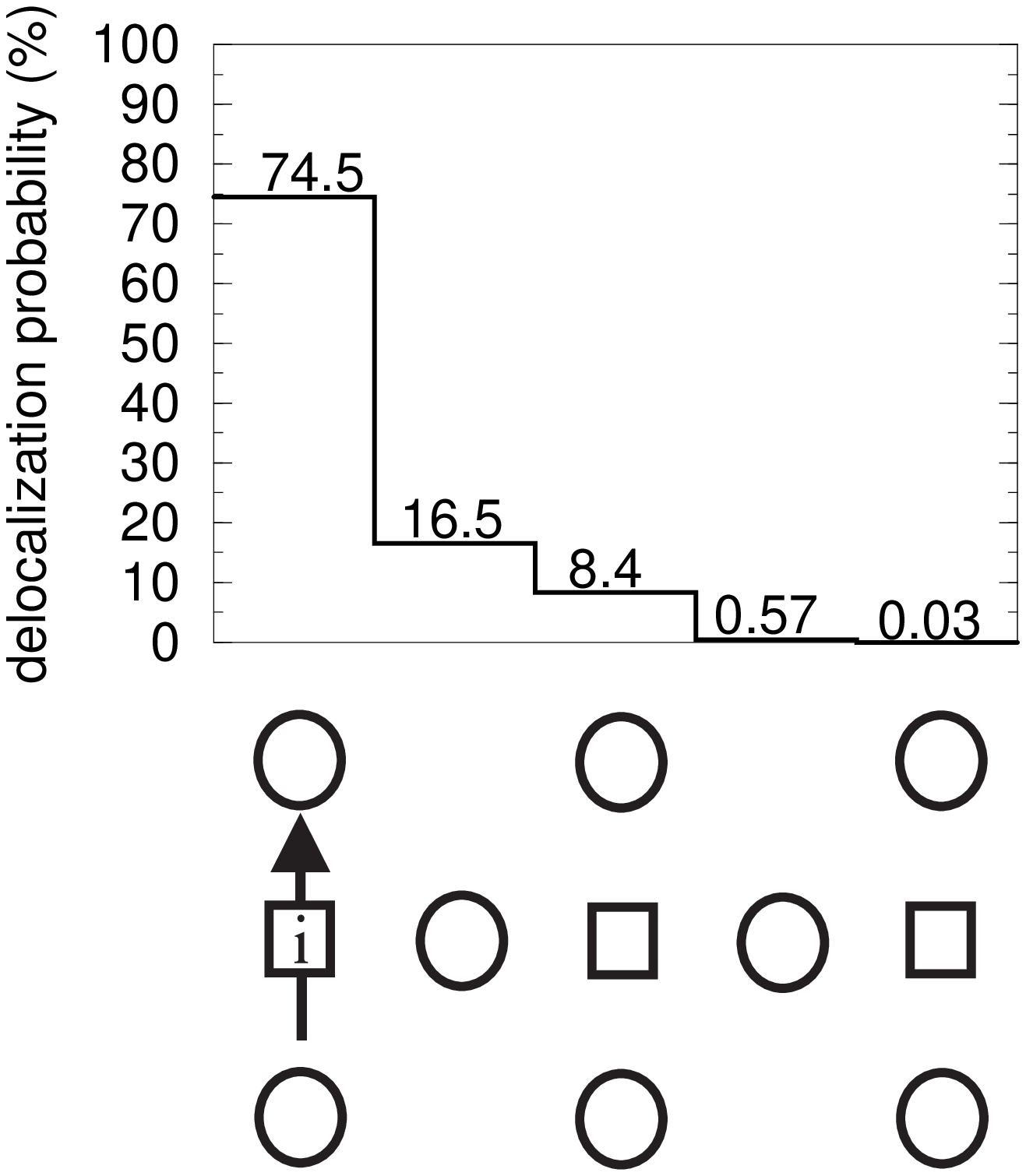}
\begin{figure}
\narrowtext
\caption{Delocalization probability as a function of fluctuation length. 
The graph shows the delocalization probability of a hole which originates 
from Cu site $i$. The probability is summed over the sites displayed 
beneath the bars, and over all equivalent final sites. The hole remains 
predominantly on its original plaquette. Delocalization beyond the 
nearest-neighbor plaquette is negligible small.
The probability has been calculated using Eq.~(\ref{groundstate4}) 
for parameter set~(\ref{parameterset}).}
\end{figure}

Figure~3 shows the delocalization probability 
$p_{\alpha }\lambda_{\alpha }^{2}/ \nu$ of a given hole summed over 
equivalent final sites as a function of fluctuation length for 
parameter set (\ref{parameterset}).
In order to demonstrate the convergence of
the results additional fluctuations have been introduced which lead beyond
the fluctuation range shown in Fig.~1. The contribution of these additional
fluctuations to the ground-state energy amounts to less than $0.1$ percent.
No significant delocalization beyond the nearest neighbor plaquette occurs.
A similar observation is made when other model-parameter values are chosen
within the range which is relevant for cuprate compounds. These results
retrospectively justify the neglect of far-reaching fluctuations and
many-body effects.
Notice, however, that the neglect of many-body effects allows for 
unphysical fluctuations which may decrease the calculated ground-state 
energy below the exact value. Thus, in contrast to an exact evaluation of 
Eq.~(\ref{expectationvalue}), our approximate solution does not guarantee
an upper limit to the exact ground-state energy.

From ground state~(\ref{groundstate4}) all expectation values are easily 
evaluated using Eq.~(\ref{expectationvalue}). The ground-state energy 
per Cu site, occupation numbers, and double occupancies of Cu and O 
sites are
\begin{eqnarray}
E_{G} &=&-4t_{pd}\lambda _{1}~\mbox{,}  \label{physicalquantities1} \\
\left\langle n_{{\rm Cu}}\right\rangle  &=&\frac{1}{\nu }\left( 1+4\lambda
_{2s}^{2}p_{2s}+4\lambda _{2d}^{2}p_{2d}\right)~\mbox{,} 
\label{physicalquantities2} \\
\left\langle d_{{\rm Cu}}\right\rangle  &=&\frac{1}{\nu }\left( 4\lambda
_{2d}^{2}p_{2d}\right)~\mbox{,}   \label{physicalquantities3} \\
\left\langle n_{{\rm O}}\right\rangle  &=&\frac{2}{\nu }\left( \lambda
_{1}^{2}+2\lambda _{3}^{2}p_{3}
+\lambda _{4}^{2}p_{4}\right)~\mbox{,}\label{physicalquantities4} \\
\left\langle d_{{\rm O}}\right\rangle  &=&\frac{1}{4}\left\langle
n_{{\rm O}}\right\rangle ^{2}~\mbox{.}  \label{physicalquantities5}
\end{eqnarray}
The number of holes is conserved, i.e. $\left\langle
n_{{\rm Cu}}\right\rangle +2\left\langle n_{{\rm O}}\right\rangle =1$. 
By comparison
with Quantum Monte Carlo calculations it will be shown in the next section 
that these results correctly reproduce the charge properties of the 
ground state. However, magnetic properties like the reduction of sublattice
magnetization due to fluctuations are only partly described. We have neglected
many-body effects in processes in which two holes simultaneously leave their
original plaquette. Therefore, no spin-flip effects are included. Thus, in 
the (Heisenberg-) limit of infinitely large $\Delta $ and $U_{d}$ ground 
state (\ref{groundstate4}) reduces to the N\'{e}el state.


\section{Discussion of the results and comparison to quantum-Monte Carlo
         simulations}

We have carried out numerical simulations of the three-band Hubbard 
model~(\ref{3-band1}), using the Projector Quantum Monte Carlo (PQMC) 
algorithm\cite{vonderLinden92}, in order to compare them with the analytical 
result, Eq.~(\ref{groundstate4}). In the PQMC approach the ground state of a 
finite cluster is projected out from a suitable trial state by applying 
the exponential operator $e^{-\beta H}$ onto the trial state in the limit 
$\beta \rightarrow \infty$. 
However, in numerical calculations only finite values of the parameter 
$\beta$ are accessible. Therefore one has to check convergence of the 
results with respect to $\beta$. Furthermore one has to account for 
possible finite size effects.

\epsfxsize=0.4\textwidth
\epsfbox{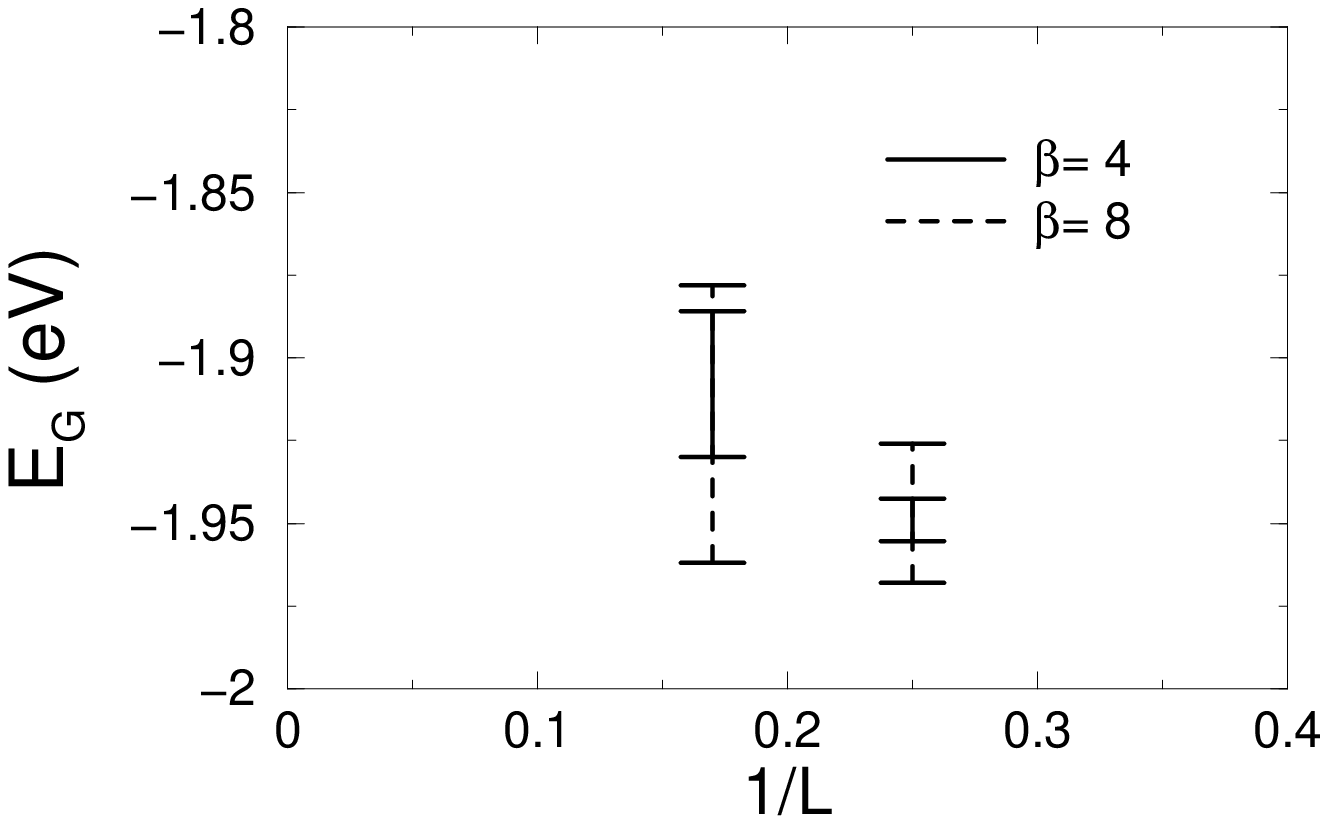}
\begin{figure}
\narrowtext
\caption{Convergence of the PQMC calculations with respect to $\beta $ and the
system size. The ground-state energy $E_{G}$ per Cu site for clusters of 
$4\times 4$ and $6\times 6$ plaquettes is shown as a function of inverse
linear system size $L$. The solid and broken error bars are the results for 
$\beta =4$ and $\beta =8$, respectively. The parameters are those of 
set~(\ref{parameterset}).}
\end{figure}

In the present study two different cluster sizes have been investigated.
Firstly, we have used a system consisting of $4\times 4$ plaquettes 
(i.e.\ $48$ sites). This is the smallest cluster which allows for periodic 
boundary conditions while still being fully two-dimensional. Secondly, 
we have studied the next largest cluster,
a system of $6\times 6$ plaquettes (i.e.\ $108$ sites).
We have used a mean-field version of the analytical 
ground state~(\ref{groundstate4}) as trial state. No sign problem occurred.

\epsfxsize=0.4\textwidth
\epsfbox{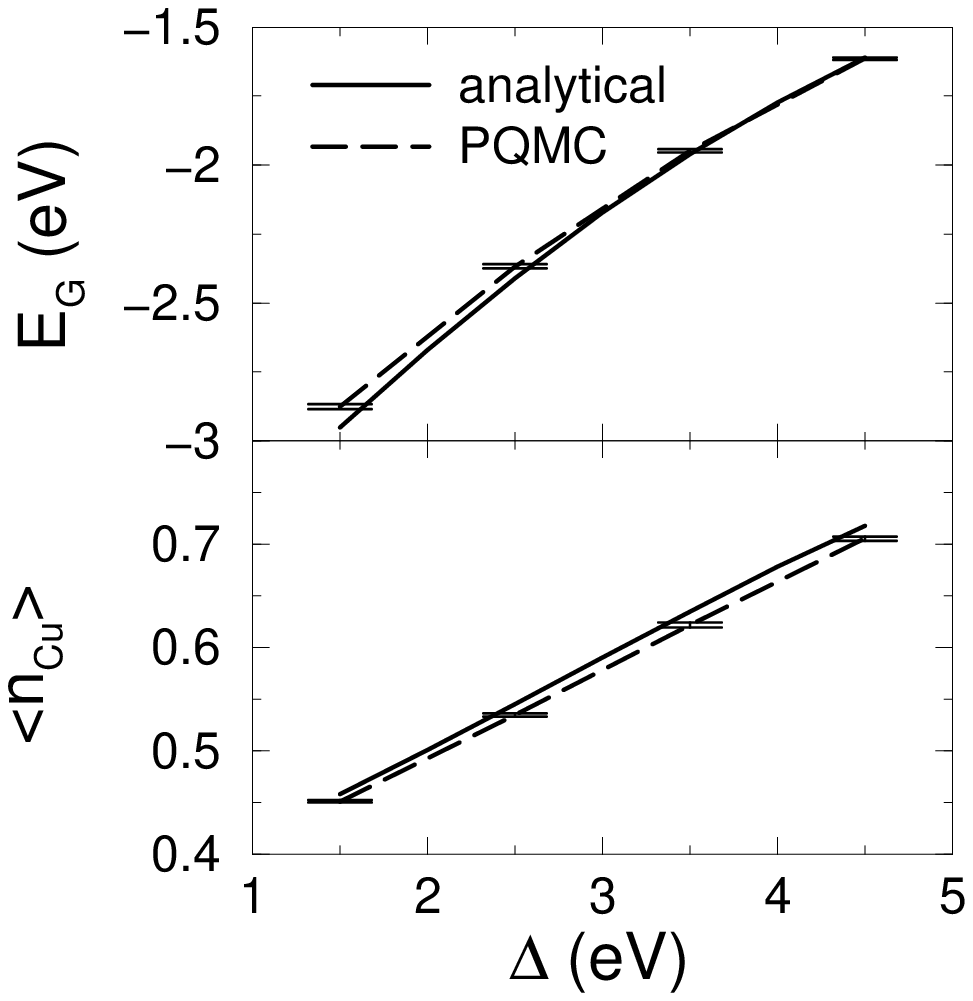}
\begin{figure}
\narrowtext
\caption{Comparison of analytically calculated ground-state energies and
Cu-occupation numbers (solid lines) with the results of PQMC simulations 
for a cluster of $4\times 4$ plaquettes (error bars connected by broken 
lines). As a function of $\Delta $ the plots show the ground-state energy 
$E_{G}$ per Cu site (upper graph), and the Cu-occupation number 
$\left\langle n_{{\rm Cu}}\right\rangle $ (lower graph). 
The other parameters are those of set~(\ref{parameterset}).}
\end{figure}

It turns out that the results obtained for the $4\times 4$ system with 
$\beta =4$ are already reasonably well converged with 
respect to both $\beta $ and system size. As shown in Fig.~4 for the case 
of the ground-state energy per Cu site the error bars for different values 
of $\beta $ overlap and the values of the $4\times 4$ system differ only
slightly from those of the $6\times 6$ system. For this reason we restrict
our simulations to a system of $4\times 4$ plaquettes with $\beta =4$ and
compare the results with the analytical approach.

In Figs.~5 and 6 the ground-state energies $E_{G}$ per Cu site
and several occupation numbers calculated using 
Eqs.~(\ref{physicalquantities1})--(\ref{physicalquantities5}) 
are compared to the results of the PQMC simulations. The number of 
holes is conserved in both approaches. Thus the O-occupation number 
$\left\langle n_{{\rm O}}\right\rangle$ is a function of the 
Cu-occupation number $\left\langle n_{{\rm Cu}}\right\rangle $, and the 
former is therefore not shown. While the values of the model parameters
are those of set (\ref{parameterset}), the charge-transfer energy 
$\Delta $ is varied covering the range of very large charge fluctuations 
$\left( \Delta =1.5\text{ eV}\right) $ to fairly small charge fluctuations 
$\left( \Delta =4.5\text{ eV}\right) $. In general there is a good agreement
between the analytical and numerical results, especially for larger values
of $\Delta $.

\epsfxsize=0.4\textwidth
\epsfbox{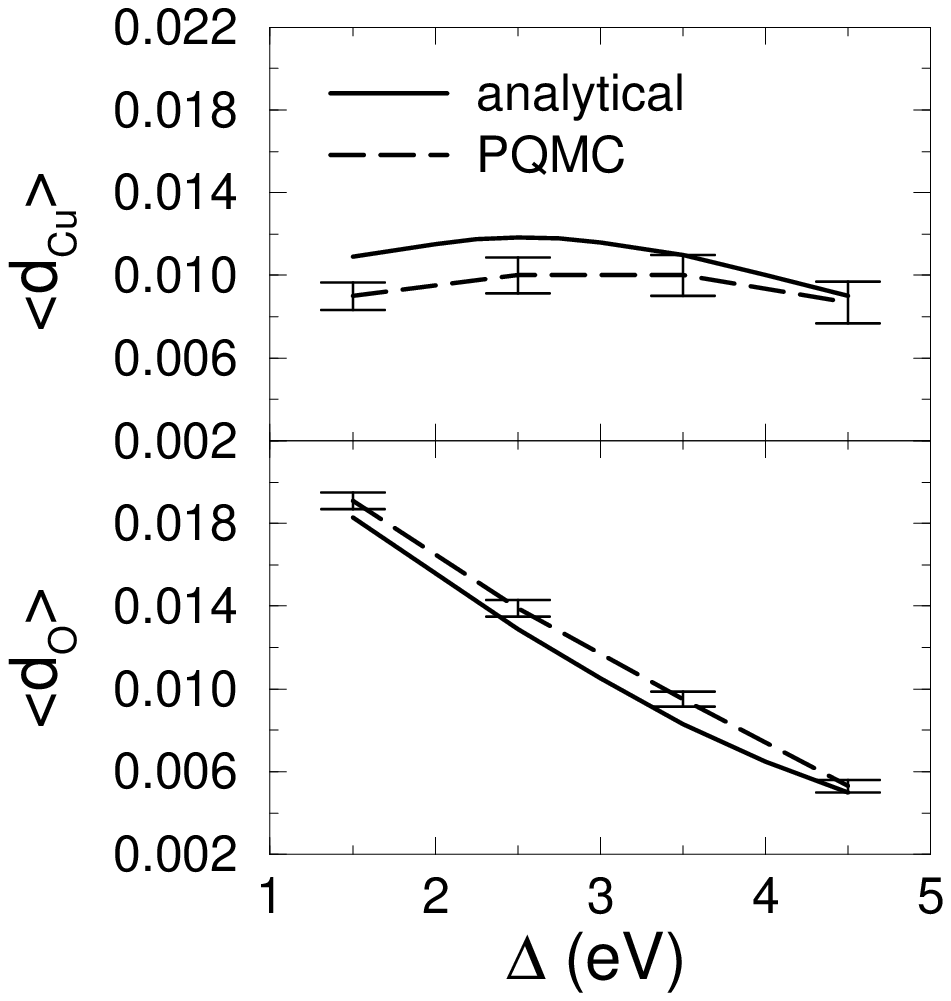}
\begin{figure}
\narrowtext
\caption{Comparison of analytically calculated double occupancies 
(solid lines) with the results of PQMC simulations for a cluster of 
$4\times 4$ plaquettes (error bars connected by broken lines). As
a function of $\Delta $ the plots show the Cu-double occupancy 
$\left\langle d_{{\rm Cu}}\right\rangle $ (upper graph), and the O-double 
occupancy $\left\langle d_{{\rm O}}\right\rangle $ (lower graph). 
The other parameters are those of set (\ref{parameterset}).}
\end{figure}

With increasing $\Delta $ both $E_{G}$ and $\left\langle n_{{\rm Cu}}
\right\rangle$ increase while the O-double occupancy $\left\langle
d_{{\rm O}}\right\rangle $ decreases. This behavior is due to the 
suppression of fluctuations for larger values of the charge-transfer 
energy. For smaller $\Delta $ the analytical value for the ground-state 
energy lies below the PQMC result. This can be explained by the neglect 
of many-body effects in Eq.~(\ref{groundstate4}) which allows for more 
unphysical fluctuations when $\Delta $ becomes smaller. These fluctuations 
also lead to values for $\left\langle n_{{\rm Cu}}\right\rangle $ which 
are slightly larger than the PQMC result. However, even for $\Delta =1.5$ eV 
the relative deviation for both $E_{G}$ and 
$\left\langle n_{{\rm Cu}}\right\rangle$ amounts to less than $3\%$.

For larger values of $\Delta $ the decrease of $\left\langle
d_{{\rm O}}\right\rangle $ with increasing $\Delta $ is about two 
times as large as the decrease of $\left\langle d_{{\rm Cu}}\right\rangle$. 
The reason for this weaker dependence of 
$\left\langle d_{{\rm Cu}}\right\rangle$ on $\Delta$ is that an increase 
in the charge-transfer energy affects $\left\langle d_{{\rm Cu}}\right\rangle$ 
only indirectly by reducing the effective Cu-Cu hopping, while - 
in contrast to $\left\langle d_{{\rm O}}\right\rangle $ - the
on-site energy of the final site is not changed. Both analytical and
numerical results show a maximum in the Cu-double occupancy $\left\langle
d_{{\rm Cu}}\right\rangle$ for intermediate values of $\Delta $. This may be
interpreted as the point where $\Delta $ is already sufficiently large to
force holes from O sites onto already occupied Cu sites but still not large
enough to suppress the effective Cu-Cu hopping.


\section{Conclusion}
Summing up, we have derived an analytical approximation, 
Eq.~(\ref{groundstate4}), for the ground state of the three-band 
Hubbard model~(\ref{3-band1}) on an infinite, half filled CuO$_{2}$ plane. 
The approach uses fluctuation operators $F_{\alpha}$ and fluctuation
strengths $\lambda_{\alpha}$ which have a clear physical interpretation.
The parameters contained in Eq.~(\ref{groundstate4}) are determined
self-consistently by solving a nonlinear system of equations. While the
approach is non-perturbative and conserves size consistency, expectation
values with the approximate ground state are still easy to evaluate.
By comparison with Projector Quantum Monte Carlo simulations we have 
demonstrated that Eq.~(\ref{groundstate4}) gives a reliable description 
of charge properties covering the range from small to very large charge 
fluctuations. Equation~(\ref{groundstate4}) can be generalized for other 
geometries. Furthermore, due to the use of fluctuation operators our 
approach provides a natural framework for the calculation of charge 
excitations, for example by using projection technique.
This will be demonstrated in a forthcoming publication.\cite{Waidacher99}


\acknowledgements
Discussions with A.~H\"ubsch and M.~Vojta are gratefully acknowledged. 
This work is supported 
by DFG through the research program of the SFB 463, Dresden.


\appendix

\section*{Relation to the Cumulant formalism}

The approach presented above can be formulated in the framework of the 
cumulant formalism.\cite{Becker88} The cumulant expectation 
value \cite{Kubo62} of a product of operators is a linear combination
of different factorizations of expectation values. For example, for two 
operators $A_{1}$ and $A_{2}$
\[
\left\langle \varphi \right| A_{1}A_{2}\left| \chi \right\rangle ^{c}=\frac{
\left\langle \varphi \right| A_{1}A_{2}\left| \chi \right\rangle }{
\left\langle \varphi \mid \chi \right\rangle }-\frac{\left\langle \varphi
\right| A_{1}\left| \chi \right\rangle \left\langle \varphi \right|
A_{2}\left| \chi \right\rangle }{\left\langle \varphi \mid \chi
\right\rangle ^{2}}~\mbox{.}
\]
Here and in the following we always assume that the states involved in a
cumulant have non-vanishing overlap, i.e. $\left\langle \varphi \mid \chi
\right\rangle \neq 0$. 
One of the attractive features of cumulants is that they preserve size
consistency.\cite{Becker88}

The cumulant formalism for the calculation of ground-state properties may
be formulated as follows. By application of an operator $\Omega $ within
the cumulant ordering an approximate ground state 
$\left| \psi _{0}\right\rangle$ can be mapped onto
the full ground state $\left| \Psi \right\rangle $ 
\begin{equation}
\left| \Psi \right\rangle ^{c}=\Omega \left| \psi _{0}\right\rangle
^{c}=\exp \left( \sum_{\alpha }\lambda _{\alpha }F_{\alpha }\right) \left| 
\psi _{0}\right\rangle ^{c}~\mbox{.} \label{cumgroundstate}
\end{equation}
The parameters $\lambda _{\alpha }$ are determined using the 
following set of equations \cite{Schork92} 
\begin{equation}
0=\left\langle \varphi \right| F_{\alpha }^{\dagger }H\Omega \left| \psi
_{0}\right\rangle ^{c}~\mbox{,}  \label{cumlambda-equations}
\end{equation}
for all $\alpha $ and for an arbitrary state $\left| \varphi \right\rangle$.
These equations follow \cite{Kladko98} from the condition that $\Omega
\left| \psi _{0}\right\rangle ^{c}$ is an eigenstate of $H$.
The exponential function in Eq.~(\ref{cumgroundstate}) should be understood 
in terms of a series expansion in which the operators $F_{\alpha}$ are 
subjected to the cumulant ordering. From Eq.~(\ref{cumgroundstate}) 
ground state properties can be calculated using 
\begin{equation}
\left\langle A\right\rangle =\left\langle \varphi \right| 
A\Omega \left| \psi _{0}\right\rangle ^{c}~\mbox{.}  
\label{cumexpectationvalue}
\end{equation}
Next we show that Eq.~(\ref{groundstate1}) and Eq.~(\ref
{lambda-equations1}) can be derived from the above equations when the
following identity \cite{Kladko98} is used 
\begin{equation}
\left\langle \varphi \right| e^{F^{\dagger }}Ae^{F}\left| \chi \right\rangle
^{c}=\left\langle e^{F}\varphi \right| A\left| e^{F}\chi \right\rangle ^{c}
~\mbox{.}\label{kladkolemma}
\end{equation}
Equation~(\ref{kladkolemma}) holds for all operators $F$ and $A$. 
It allows to remove the exponential functions from 
the cumulant ordering and apply them directly onto the states.
Using Eq.~(\ref{kladkolemma}), Eqs.~(\ref{groundstate1}) and
(\ref{cumgroundstate}) can be directly transformed into each other. 
In Eq.~(\ref{cumlambda-equations}), on the other hand, we choose 
$\left| \varphi\right\rangle =\Omega \left| \psi _{0}\right\rangle $ 
and use Eq.~(\ref{kladkolemma}) to obtain 
\[
0=\left\langle \Omega \psi _{0}\right| F_{\alpha }^{\dagger }H\left| \Omega
\psi _{0}\right\rangle ^{c}~\mbox{.}
\]
If $\left| \Omega \psi _{0}\right\rangle $ is an eigenstate of $H$ this
equation is equivalent to Eq.~(\ref{lambda-equations1}).

We conclude by pointing out some of the advantages of 
Eqs.~(\ref{groundstate1}) and (\ref{lambda-equations1}) 
as compared to other possible approaches
within the framework of the cumulant formalism. First, the fact that the
exponential operator $\Omega $ has been transferred onto the state
$\left| \psi _{0}\right\rangle$ amounts
to a summation of all orders in the $\lambda _{\alpha }$. Thus, our approach
is non-perturbative and avoids (possibly divergent) series expansions of
expressions like Eq.~(\ref{plaqncu}). The divergence of these series 
for $\lambda _{1}\geq 0.5$ is equivalent to the violation of 
condition\cite{Schork94} 
$\left|\left\langle \psi _{0}\mid \Psi \right\rangle \right| ^{2}>1/2$. 
Consequently, when $\lambda _{1}\geq 0.5$, within the
cumulant ordering no operator $\Omega $ exists for state $\left| \psi
_{0}\right\rangle $. These difficulties do not occur when $\Omega $ has been
removed from the cumulant ordering. Second, since in Eq.(\ref
{lambda-equations1}) the full ground state $\left| \Psi \right\rangle $ 
appears as both bra and ket vector more fluctuations are taken into account.
If, for example, Eq.~(\ref{cumlambda-equations}) with $\left| \varphi
\right\rangle =\left| \psi _{0}\right\rangle $ is used instead of Eq.~(\ref
{lambda-equations1}), one always obtains a vanishing value for $\lambda _{2s}
$.


\end{multicols}

\end{document}